\documentclass[aps,prb,preprint]{revtex4}
\usepackage{graphicx}

\def\TMC{T_{\mathrm{MC}}}
\def\eIS{e_{\mathrm{IS}}}
\def\oBP{\omega_{\mathrm{BP}}}
\def\ec{e_{\mathrm{c}}}

\begin{document}

\title{Phonons in supercooled liquids: a possible explanation for the
Boson Peak}

\author{T.~S.~Grigera$^\dagger$, V.~Mart\'\i{}n-Mayor$^\#$,
G.~Parisi$^*$ \& P.~Verrocchio$^\#$}

\affiliation{$\dagger$ Centro di Studi e Ricerche ``Enrico Fermi'', via
 Panisperna 89/A, I-00184 Roma, Italy \\ $*$ Dipartimento di Fisica ,
 Sezione INFN, SMC and INFM unit\`a di Roma 1, Universit\`a di Roma La
 Sapienza, Piazzale Aldo Moro 2, I-00185 Rome, Italy\\ $\#$
 Departamento de F\'isica Te\'orica I, Universidad Complutense de
 Madrid, Avenida Complutense s/n, 28040 Madrid, Spain }

\date{\today}

\maketitle

{\bf Glasses\cite{Angell95,DeBenedetti01} are amorphous solids, in the
sense that they display elastic behaviour. In crystals, elasticity is
associated with phonons, quantized sound-wave excitations. Phonon-like
excitations exist also in glasses at very high frequencies (THz), and
they remarkably persist into the supercooled liquid\cite{Sette98}.  A
universal feature of these amorphous systems is the Boson peak: the
vibrational density of states $g(\omega)$ has an excess over the Debye
(squared frequency) law, seen as a peak in $g(\omega)/\omega^2$. We
claim that this peak is the signature of a phase transition in the
space of the stationary points of the energy, from a minima-dominated
phase (with phonons) at low energy to a saddle-point dominated
phase\cite{Angelani00,Broderix00,Grigera02} (without phonons).  Here,
by studying the spectra of inherent structures\cite{Stillinger95}
(local minima of the potential energy), we show that this is the case
in a realistic glass model: the Boson peak moves to lower frequencies
on approaching the phonon-saddle transition and its height diverges at
the critical point.  The numerical results agree with Euclidean Random
Matrix Theory\cite{Mezard99} predictions on the existence of a sharp
phase transition\cite{JPHYS02} between an amorphous elastic phase and
a phonon-free one. }

Vibrational excitations of glasses in the THz region, and the related
vibrational density of states (VDOS), play a crucial role in their
thermodynamic properties \cite{Phillips}. Recent results suggest that
the VDOS is determined by the properties of the potential energy
landscape \cite{Stillinger95} (PEL), a crucial tool to understand the
slow dynamics of supercooled liquids \cite{DeBenedetti01}.  Indeed,
the VDOS and the dynamic structure factor can be qualitatively
reproduced from the (harmonic) vibrational spectrum obtained from the
diagonalization of the Hessian matrix of the potential energy
evaluated at the minima (or inherent structures [IS]) of a
Lennard-Jones system \cite{Ruocco00}.  The same numerical procedure
led to quantitative agreement with inelastic X-ray scattering
experiments in amorphous silica (O.~Pilla et al., preprint
condmat/0209519). Here we will show that a Boson peak must be present
in the VDOS as a consequence of the PEL geometry,
and make quantitative predictions which can be checked fastly
quenching supercooled liquids below their glass temperature.

The high-frequency ($0.1$--$10$ THz) excitations have been
experimentally shown to have linear dispersion relations
\cite{Sette98,Masciovecchio96,Matic01,Masciovecchio97,Benassi96,Fioretto99,Pilla00}
in the mesoscopic momentum region ($\sim 1$--$10$ nm$^{-1}$).
Although clearly not plane waves, they are highly reminiscent of
phonons because they propagate with the speed of sound and because the
VDOS, $g(\omega)$, is Debye-like ($g(\omega) \propto \omega^2$) at low
enough frequency.  These excitations have in fact been dubbed
``high-frequency sound'' and a coherent theoretical picture of their
properties was obtained \cite{JPHYS02,Martin01,Grigera01} using
Euclidean Random Matrix Theory~\cite{Mezard99} (ERMT).

However, there is more to the low frequency VDOS of glasses than the
Debye law.  A characteristic feature is that the VDOS departs from the
Debye form, displaying an excess of states, which has been named
``Boson peak'' (BP). As observed in many materials
\cite{Sette98,Masciovecchio96,Matic01,Masciovecchio97,Benassi96,Fioretto99,Pilla00},
the BP is in a region of frequencies where the dispersion relation for
phonons is still linear ($\omega \sim$ a few THz).  Several
non-equivalent ways of experimentally defining the BP have been
introduced.  One approach defines it as a peak in Raman-scattering
data, another as a peak in the difference between the VDOS of the
glass and the corresponding crystal. Other authors extract $g(\omega)$
from their neutron scattering data and look for a peak in
$g(\omega)/\omega^2$.  This is the definition we adopt, since we
believe it unveils universal properties of glasses.

The origin of the Boson peak can be understood if we consider the
ensemble of generalized inherent structures (GIS): for each
equilibrium configuration the associated GIS is the nearest stationary
point of the Hamiltonian.  If we start from an equilibrium
configuration at low temperature, the GIS is a local minimum, and it
coincides with the more frequently used IS (i.e.\ the nearest minimum
of the Hamiltonian). On the contrary, if we start from high
temperature, the GISs are saddle points. In the GIS ensemble there is
a sharp phase transition separating these two regimes.  It takes place
in glass-forming liquids~\cite{Angelani00,Broderix00,Grigera02} at the
Mode Coupling temperature~\cite{Goetze92,Kob95} ($\TMC$), above which
liquid diffusion is no longer ruled by rare ``activated'' jumps
between ISs but by the motion along the unstable directions of
saddles.  Phonons are present in the spectrum of the VDOS in the low
temperature phase (IS dominated) but are absent in the saddle phase
(see ref.~\onlinecite{Goetze00} for a different approach).

The key point, following both from analytic computations in soluble
models \cite{Cavagna00} and simulations
\cite{Angelani00,Broderix00,Grigera02}, is that the minima obtained
starting from configurations below $\TMC$ and the saddles obtained
starting above $\TMC$ join smoothly at $\TMC$. Thus we can study GIS
as a single ensemble, parametrized by the initial temperature or the
final energy. We expect that the GIS ensemble belong to a large
universality class.

This transition from the phonon phase to the saddle phase is a quite
general phenomenon that it is not restricted to glasses. It has been
studied in the framework of the Euclidean Random Matrix Theory (ERMT)
\cite{Mezard99}, showing \cite{JPHYS02} that close to this transition,
a BP is present in the {\em phonon} phase, whose position shifts to
lower frequencies on approaching the transition point, while its
height grows without bound.  The BP actually signals a crossover at
frequency $\oBP$ between a (phonon dominated) $\omega^2$ scaling of
$g(\omega)$ to an $\omega^{\gamma}$ scaling ($\gamma<2$) that is
present at the phase transition point.  More precisely, at frequencies
small respect to the Debye frequency, the VDOS should satisfy the
scaling law
\begin{equation}
    g(\omega,\Delta)=\omega^{\gamma} h(\omega \Delta^{-\rho}),
\label{sc1}
\end{equation} where $\Delta$ is the distance from the critical point
and depends on the actual control parameter (pressure, temperature,
etc.) and $h(x)$ is such that $h(x) \sim x^{2-\gamma}$ for $x \ll 1$
and $h(x) \sim$ constant for $x \gg 1$. This scaling law implies that
\begin{equation}
\oBP \propto \Delta^\rho, \qquad 
 \frac{g(\oBP)}{\oBP^2}\propto \Delta^{-\beta} , 
 \label{scaling}
\end{equation}
where $\beta=\rho (2-\gamma)$. 
Under the resummation of a given class of diagrams, ERMT predicts
\cite{JPHYS02} that $\rho=1$, $\gamma=3/2$ and $\beta=1/2$.  Of course
the actual (universal) values of these critical exponents in three
dimension may differ slightly from the values found in this
approximation.

We now analyze numerically the existence of such a transition by
studying the soft sphere binary mixture, a model of a fragile
glass~\cite{Bernu87}. We simulate this system with the Swap Monte
Carlo algorithm \cite{GrigeraParisi01}, and compute the VDOS of the
ISs obtained starting from equilibrium configurations at temperatures
below $\TMC$.
Except were stated, we used
2048 particles, and runs were followed until the energy of the IS,
$\eIS$, reached a stationary value. 
The predictions above agree with the numerical data, taking $\Delta =
\ec-\eIS$ ($\eIS$ being the energy of the ISs and $\ec$ the critical
value).

Given the numerical results of Kob et al.~\cite{Kob00}, one expects
that $g(\omega)$ depend only on $\eIS$, and thus that $\eIS$ be the
relevant control parameter.  To check this, we compare the VDOS of two
systems at different temperatures (one in equilibrium, the other not)
but with the same $\eIS$. The equilibrated system has 2048 particles,
while the other has 20000 and is in the regime where $\eIS$ has not
yet stabilized after a quench (Fig. 1b). As Fig. 1a shows for a
representative case, the VDOS obtained at the same $\eIS$ for the two
systems coincides, confirming that $\eIS$ is the control parameter. We
can thus proceed to investigate the scaling laws (\ref{scaling}). 

The VDOS at three representative temperatures is shown in Fig.~2.  For
temperatures up to $T=0.69 \, \TMC$, a Debye behaviour ($g(\omega)\propto
\omega^2$) is found for the lowest frequencies that can be resolved in
a system of this size. In the $g(\omega)/\omega^2$ plot (Fig.~2a), a
peak is clearly identified, which is seen to grow in height and shift
to lower frequency on rising the temperature. The peak shifts to
frequencies below our resolution for $T>0.69 \, \TMC$, but we can identify a
Debye region up to $T=0.83 \, \TMC$. Above this temperature the Debye region
is not well defined, and it disappears completely for $T \ge
0.9 \, \TMC$. At these temperatures, $g(\omega)/\omega^2$ seems to diverge
(at least within the frequency range that we can
resolve). Experimentally, this behaviour may be difficult to observe
due to the presence of the elastic peak, but B$_2$O$_3$ data
\cite{Engberg99} seem to support this prediction. For frequencies
immediately above $\oBP$, the VDOS scales as $\omega^{\gamma}$, with
$\gamma\approx 1.5$ (Fig.~2b). Note that these changes of the VDOS
with temperature go practically unnoticed if one plots $g(\omega)$
directly (Fig.~2c). In particular, the BP is completely unrelated to
the first maximum of $g(\omega)$, whose position is insensitive to
temperature changes.

Since we know that $\eIS$ decreases slowly with time after a quench
\cite{Kob00} (see also Fig.~1b), the above results enable us to make
predictions about the {\em ageing} of the VDOS, and in particular of
the BP. With increasing time the system moves farther from the
critical point, and thus the BP should decrease in height and shift to
higher frequencies.  Moreover, at a given frequency (below the
asymptotic $\oBP$) $g(\omega)$ should decrease, since it will be of
order $\omega^\gamma$ at short times and of order $\omega^2$ at very
long times.  Similarly, one should expect a cooling rate dependence of
the shape of the BP: the slower the cooling, the lower the asymptotic
$\eIS$ and thus a larger $\oBP$ and less pronounced BP.  An effect of
this kind has been observed in As$_3$S$_3$ \cite{Mamedov98}, although
here it is difficult to disentangle the physical effect from the
chemical changes of the sample due to dissociation. To our knowledge,
a clear-cut experimental observation of these effects has not been
reported.

Using all the spectra for which the peak position can be clearly
identified, we find that the relationship between $\oBP$ and the
energy of the IS is linear (Fig. 3a).  The energy at which $\oBP$
becomes zero, $\ec$, is found from a linear fit as $\ec=1.74 \,
\epsilon$ ($\epsilon$ is the energy scale). As for the height of the
peak, the points up to $T=0.83 \, \TMC$ are compatible with a
power-law divergence of the form (\ref{scaling}) (Fig.~3b). Fixing
$\ec$ at the value $1.74 \, \epsilon$ arising from the linear fit of
$\oBP$ {\sl vs.} $\eIS$, a power-law fit yields an exponent
$\beta=0.40(15)$, while if one fixes the exponent $\beta=1/2$, then the
critical value is obtained as $\ec=1.752(2) \, \epsilon$. Thus the
numerical data are compatible with the theoretically predicted
scaling, although we have not been able to work close to the critical
point, and thus cannot get a great accuracy on the critical exponents
or the critical point.

The present discussion applies to experiments as long as one is in the
regime where the inverse frequency is much larger than the structural
relaxation time, when the harmonic approximation makes sense
\cite{Horbach99}. Thus, while the BP observed in liquids can be
understood by the present considerations, one should not expect to
actually arrive close to the critical point in equilibrium
measurements. Rather, one should generate a glass by hyperquenching
(see e.g.\ talk by C.~A.~Angell in
http://www.df.unipi.it/workshop/oral/OL6.pdf): for this we predict
that the BP measured at low T should obey the scaling~(\ref{scaling}),
the control parameter being the fictive temperature,
which in our instantaneous quench is just the temperature at which the
hyperquench starts, but in experiments involves also the cooling rate.
In this way one should also alleviate the problem of the swamping by
the quasielastic peak.

The significance of the value $\ec=1.74(1) \, \epsilon$ is
twofold. First, it is the value of $\eIS$ that corresponds, in
equilibrium, to $T=\TMC$, the Mode-Coupling critical temperature
(which we have independently determined as in ref.~\onlinecite{Kob95},
using a standard Monte Carlo without swap). Second, it coincides
within errors with the threshold energy below which the ratio of the
number of saddle points to the number of minima vanishes exponentially
with the number of particles~\cite{Grigera02} (our numerical accuracy
on $\ec$ is similar to that of~\onlinecite{Grigera02}).  Thus we find
a link between the onset of slow dynamics, the appearance of
phonon-like excitations and a geometrical phase transition of the PEL
in fragile glasses. At variance with previous PEL studies, the clear
link between the PEL geometrical transition and high-frequency
dynamics allows it to be studied experimentally by following the
evolution of the BP.

In summary, we have studied numerically the VDOS of a fragile
glass-former liquid undergoing its Mode-Coupling transition. We have
found that in the ensemble of the stationary points near the
equilibrium configurations (GIS) there is a transition from a
mechanically unstable phase where saddles are present to a stable
phase where phonon-like excitations appear. The scaling laws predicted
by ERMT are compatible with the numerical results we have obtained for
the soft sphere model.  This approach is equivalent to other
microscopic descriptions of the glass transition based on the geometry
of the potential energy landscape (PEL)
\cite{Angelani00,Broderix00,Grigera02,LaNave02}. But this point of
view emphasizes quantities accessible to experiment, like the VDOS,
and proposes an interpretation of an universal feature of glasses, the
Boson peak. The BP appears in the phonon phase, and it is a signature
of a cross-over from a phonon-dominated spectrum with a Debye
$\omega^2$ scaling to an $\omega^{\gamma}$, $\gamma\approx 1.5$
spectrum, resulting from the hybridization of acoustic modes with
high-energy modes that soften upon approaching the saddle-phonon
transition \cite{JPHYS02}. The ERMT predictions (eqs.~\ref{sc1}
and~\ref{scaling}) could be checked experimentally in hyperquenching
experiments, where
the Boson peak should strongly depend on the fictive temperature.  We
expect that the saddle-phonon transition point of view will be able to
bridge the realms of experiment and numerical studies of the PEL,
allowing to test many geometrical ideas and to use the PEL insights in
the detailed analysis of the experimental glass transition.

\section*{Acknowledgements}

We thank O.~Pilla, G.~Ruocco and G.~Viliani for helpful
discussions. We are grateful to the RTN3 collaboration for CPU time in
their cluster.  VMM was supported in part by European Commission and
OCYT (Spain).

\newpage
\begin{figure}
\includegraphics[angle=90,width=\columnwidth]{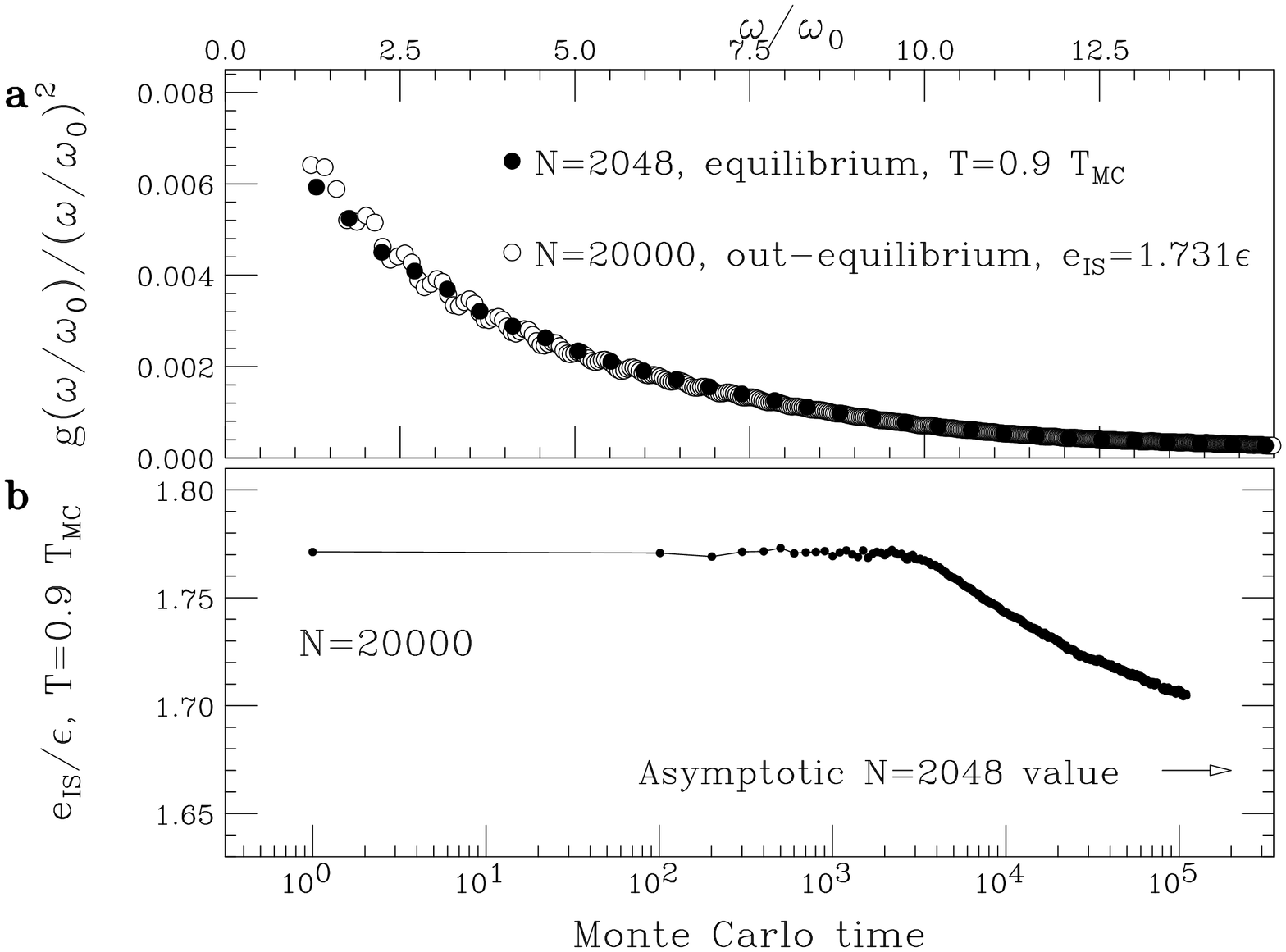}
\end{figure}

\begin{figure}
\caption{The VDOS $g(\omega)$ at low frequencies depends only on the
energy of the inherent structure, $\eIS$. {\bf a.}
$g(\omega)/\omega^2$ for an equilibrated 2048 particle system at $T =
0.9 \, \TMC$ compared to that obtained from ten inherent structures of
the same $\eIS$ from a 20000-particle system out of equilibrium.  The
larger system was quenched from infinite temperature to $T = 0.54 \,
\TMC$ and the inherent structures obtained by minimizing instantaneous
configurations between $1.9\cdot 10^4$ and $2\cdot 10^4$ Swap Monte
Carlo steps. Similar results are obtained at all the temperatures for
which comparable $\eIS$ are obtained in the quench. IS were found by
minimizing instantaneous configurations with a conjugate gradient
algorithm. The Hessian was diagonalized with standard library routines
in the case of the small system, and with the method of moments for
the big one.  Each VDOS was obtained from at least 200 IS as the
histogram of the square root of the eigenvalues.  {\bf b} Monte Carlo
history of $\eIS$ in logarithmic time-scale, for the system of 20000
particles, suddenly quenched from infinite temperature to $T=0.54 \,
\TMC$.  The soft-sphere binary mixture is made of 50\% of particles of
type A and 50\% of type B, both with the same mass. The interaction
potential is $V_{\alpha \beta}(r)= \epsilon [(\sigma_\alpha +
\sigma_\beta)/r]^{12} + C_{\alpha \beta}$ ($\alpha, \beta \in
\{A,B\}$), with $\sigma$'s fixed by the conditions $\sigma_B=1.2
\sigma_A$, and $(2\sigma_A)^3 + 2 (\sigma_A+\sigma_B)^3 +
(2\sigma_B)^3=4\sigma_0$.  A smooth cutoff is imposed at $r_c=\sqrt{3}
\sigma_0$: for $r_c \le r \le a$, we set $V_{\alpha \beta}=B_{\alpha
\beta} (a-r)^3$, and $V_{\alpha \beta}=0$ for $r>a$.  In the results,
lenghts are given in units of $\sigma_0$, energies in units of
$\epsilon$ and frequencies in units of $\omega_0 \equiv (m \sigma^2_0
/ \epsilon)^{-1/2}$. Using Argon parameters ($\sigma_0=3.4 \AA$,
$\epsilon=120\,$K$ k_B$, $m=39.96 \,$ u.a.m.) the frequency unit
$\omega_0$ is $0.46 \, Thz$ and the Mode Coupling temperature $\TMC$
is $26.4 K$.  All runs are at constant volume, with particle density
$\rho= \sigma_0^{-3}$.  For temperatures down to $T=0.9 \,\TMC$
thermalization was achieved, as checked by ensuring that the
equilibrium fluctuation-dissipation ratio is reached.  For $T < 0.9 \,
\TMC$, the run was followed until $\eIS$ reached stationarity.}
\end{figure}

\begin{figure}
\includegraphics[angle=90,width=\columnwidth]{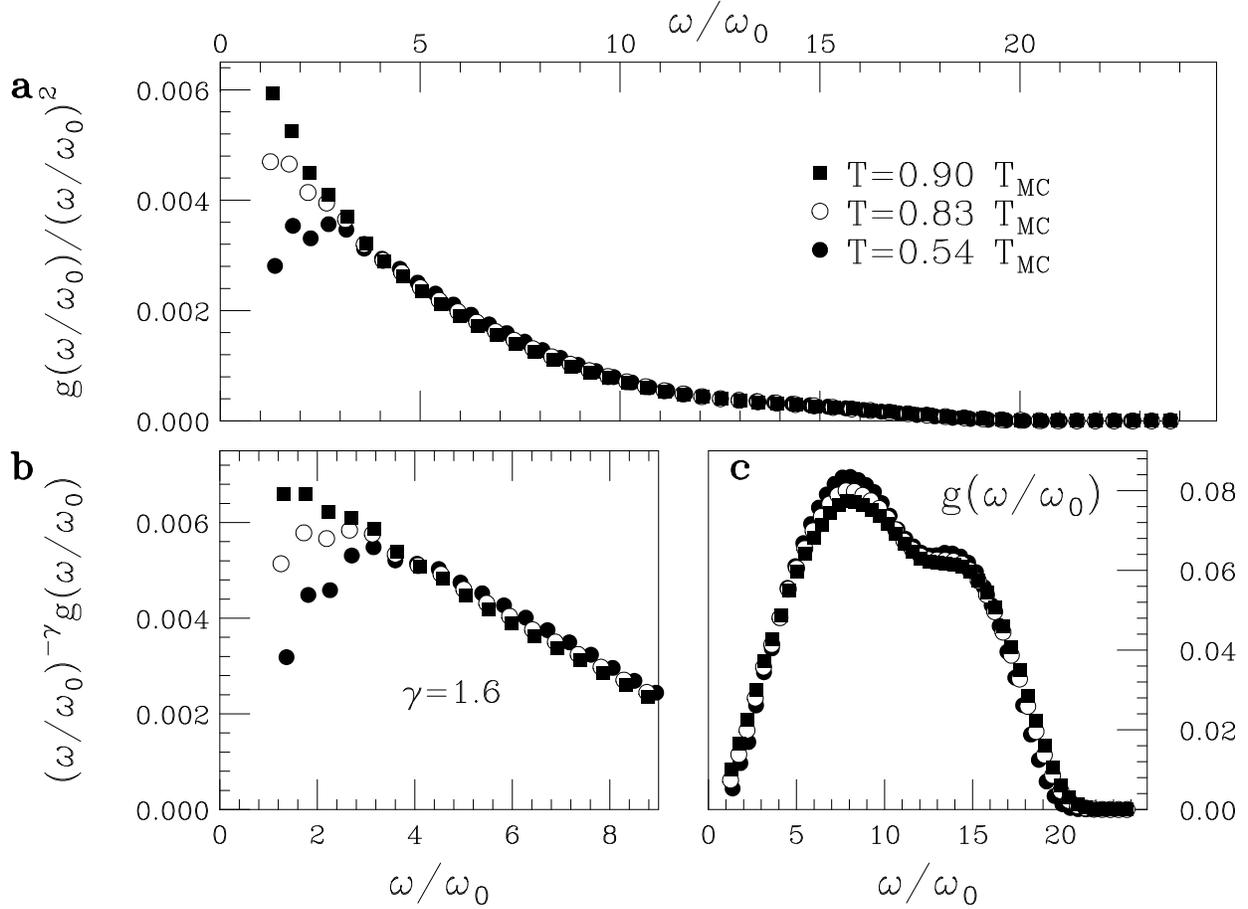}
\caption{Vibrational density of states $g(\omega)$ of the soft-sphere
binary mixture at three representative temperatures (the full set of
temperatures was $T/ \TMC = 0.9, 0.83, 0.78, 0.69, 0.61, 0.54, 0.49$,
frequencies are given in units of $\omega_o$). {\bf a.} The
$g(\omega)/\omega^2$ plot shows a Boson peak at low temperature (shown
$T=0.54 \, \TMC$). When temperature is risen, the peak shifts to lower
frequencies and grows without bound. {\bf b.} The
divergence seen in {\bf a} is due to an $\omega^\gamma$ (with
$\gamma<2$) scaling of
$g(\omega)$. At temperatures where the Boson peak is still seen,
a crossover between $\omega^2$ and $\omega^\gamma$ scaling can be noted
comparing {\bf a} and {\bf b.} While $\gamma$ cannot be extracted very
accurately from the data, it is compatible with $\gamma=2-\beta/\rho
\approx 1.6$, with $\beta$ and $\rho$ taken from the fits of Fig.~3.
{\bf c.} The high frequency features of $g(\omega)$ do not change
drastically with temperature. }
\end{figure}

\begin{figure}
\includegraphics[angle=90,width=\columnwidth]{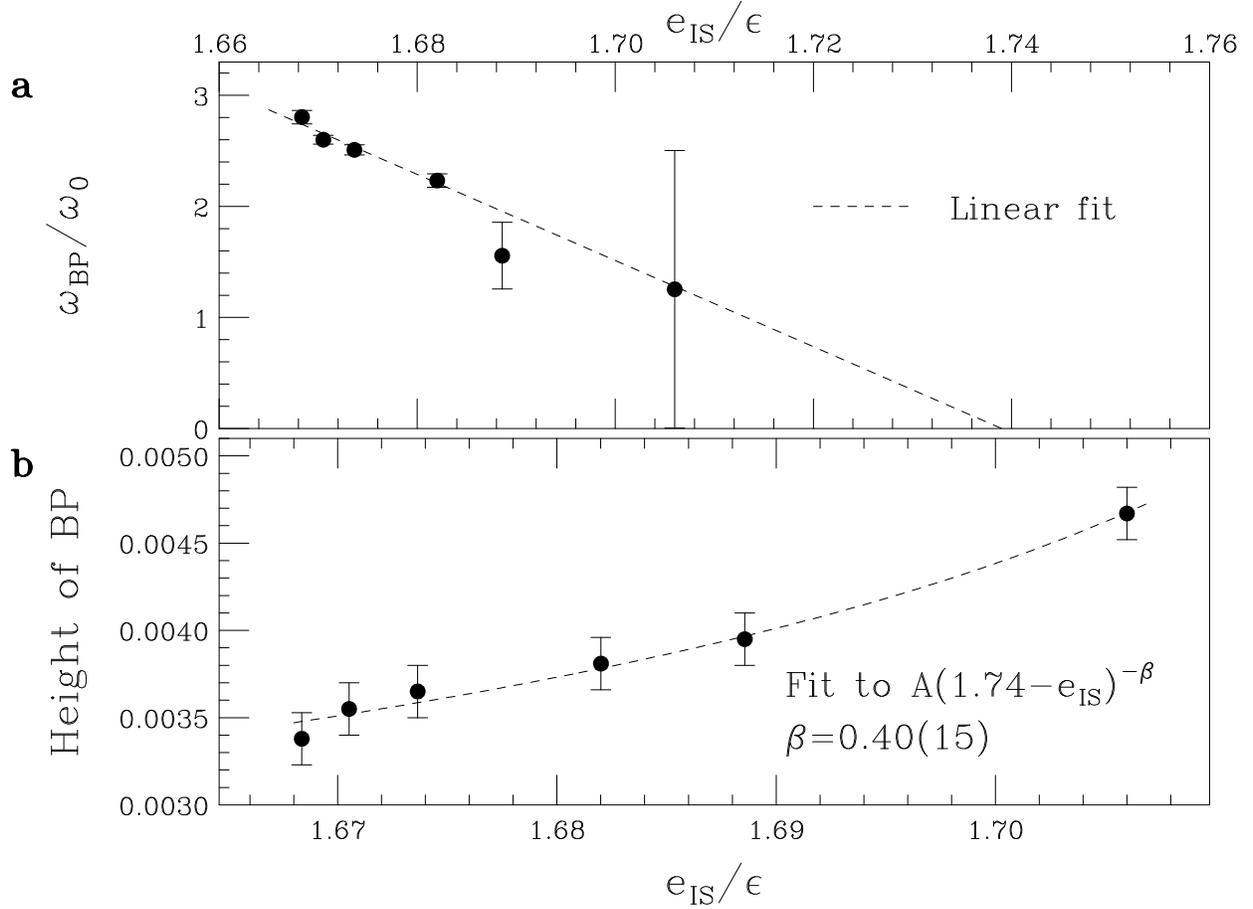}
\caption{Scaling of the position, $\oBP$, and height of the Boson peak
near the saddle-phonon transition (energies and frequencies in units
of $\epsilon$ and $\omega_0$ respectively). {\bf a.} $\oBP$ is linear
in the control parameter, in this case the energy of the inherent
structures $\eIS$. Extrapolating with a linear fit, $\oBP$ goes to 0
at $e_{IS}=\ec=1.74(1) \, \epsilon$. {\bf b.} The height of the Boson peak
diverges as a power law (the leftmost point has been left out of the
fit). Height and position of the BP were obtained by fitting a
parabola to the peak of $g(\omega)/\omega^2$ at $T/\TMC =0.49, 0.54, 0.61,
0.69, 0.78, 0.83$. The corresponding $\eIS/ \epsilon$ are $1.668, 1.671,
1.674, 1.682, 1.689, 1.707$.}
\end{figure}

\end{document}